\documentclass[a4paper,11pt]{article}

\usepackage{contribution}

\newcommand{\weblink}[2][]{%
    \ifthenelse{\equal{#1}{}}%
    {\textnormal{\url{#2}}}%
    {\textnormal{\href{#2}{#1}}}%
}

\newcommand{\acknowledgements}[1]{%
  \bigskip\bigskip
  \textsf{\textbf{\Large Acknowledgements}} \\[2ex]
  {#1}
  \bigskip
}

\def\beq{\begin{equation}}
\def\eeq#1{\label{#1}\end{equation}}
\def\eeqn{\end{equation}}

\def\beqa{\begin{eqnarray}}
\def\eeqa#1{\label{#1}\end{eqnarray}}
\def\eeqan{\end{eqnarray}}

\let\bar=\overbar

\def\Dslash{\not{\hbox{\kern-4pt $D$}}}
\def\dslash{\not{\hbox{\kern-2pt $\del$}}}

\def\msb{{\bar{\ssstyle M \kern -1pt S}}}

\newcommand{\contribution}[7][]{%
  \clearpage
  \thispagestyle{plain}
  \ifthenelse{\equal{#1}{}}
  {\hypersetup{pdftitle={#2}}}
  {\hypersetup{pdftitle={#1}}}
  \hypersetup{pdfauthor={{#3} {#4}}}
  {\centering\normalfont\LARGE\bfseries\sffamily #2 \par\nobreak}
  \lhead{}
  \chead{%
    \textit{\footnotesize XIV International Conference on Hadron Spectroscopy
      (\weblink[\textit{hadron2011}]{http://www.hadron2011.de}), 13-17 June 2011, Munich, Germany}%
  }
  \rhead{}
  \bigskip
  \begin{center}
    {#3} {#4}\ifthenelse{\equal{#6}{}}{}{\footnote{\weblink[#6]{mailto:#6}}}
    \ifthenelse{\equal{#7}{}}{}{#7} \\
    \textit{#5}
  \end{center}
  \bigskip
}

\renewcommand{\abstract}[1]{%
  \begin{center}
    \begin{minipage}{0.85\textwidth}
      \begin{footnotesize}
        #1
      \end{footnotesize}
    \end{minipage}
  \end{center}
  \bigskip
}

\begin{document}

%
%
%
%
%
{  

%

\contribution[Photoproduction of $\eta^{\prime}$ Mesons from Nuclei]  
{Photoproduction of $\eta^{\prime}$ Mesons from Nuclei}  
{Mariana}{Nanova}  
{II. Physikalisches Institut \\
  Universit\"at Giessen,\\
  D-35392 Giessen, GERMANY}  
{Mariana.Nanova@physik.uni-giessen.de}  
{on behalf of the CBELSA/TAPS Collaboration}  
%

\abstract{%
The photoproduction of $\eta^{'}$-mesons from different nuclei has been measured using the  Crystal Barrel(CB)/TAPS detector system at the ELSA accelerator facility in Bonn.
Recent results on the in-medium properties of the $\eta^{'}$-meson, derived from the transparency ratio measurements, are presented. The absorption of the $\eta^{'}$-meson in nuclear matter is compared to the properties of the other mesons ($\eta$ and $\omega$).
}
%

\section{Introduction}
As we know from hadron physics, the light pseudoscalar mesons ($\pi$, $K$, $\eta$) are the Nambu-Goldstone bosons associated with the spontaneous breaking of the QCD chiral symmetry. Introducing the current quark masses these mesons together with the heavier $\eta^{'}$(958) meson show a mass spectrum which is believed to be explained by the explicit flavor $SU(3)$ breaking and the axial $U_{A}(1)$ anomaly. Recently, there have been several important developments in the study of the spontaneous breaking of chiral symmetry and its partial restoration at finite density. 
Theoretical studies predict a possible mass shift of the $\eta^{'}$-meson at finite density and the formation of bound states ~\cite{nagahiro}. Up to the last  two years $\eta^{'}$ photoproduction was not much explored, but with new generation experiments results on differential and total cross sections of $\eta^{'}$ photoproduction on the proton ~\cite{crede} and the deuteron ~\cite{igal} have now become available. \\ 
The investigation of $\eta^{\prime}$-meson production in photon induced reactions on solid target provides information on in-medium properties of the meson. The in-medium width of the $\eta^{\prime}$-meson can be extracted from the attenuation of the meson flux deduced from a transparency ratio measurement. However, $\eta^{\prime}$-meson will decay outside of the nucleus because of their long lifetime and thus their in-medium mass is not directly accessible experimentally. Consequently, an experimental observation of  a possible mass shift of  the $\eta^{\prime}$-meson in-medium is extremly difficult. In the present work, we concentrate on the transparency ratio measurement of  the $\eta^{\prime}$-meson and disccuss its absorption properties.

\section{Data Analysis}
The experiment has been performed at the ELSA facility in Bonn ~\cite{Husmann,Hillert} using the Crystal Barrel(CB) ~\cite{aker92}  and TAPS~\cite{Novotny1}  detector system (Fig. ~\ref{fig:exp} left). The combined Crystal Barrel/TAPS detector covered 99\% of the full 4$\pi$ solid angle. The high granularity of the system makes it very well 
suited for the detection of multi-photon final states.  Tagged photon beams of energy up to 2.6 GeV were produced via bremsstrahlung and impinged on a solid target. For the measurements targets of $C$, $Ca$, $Nb$ and $Pb$ were used. A more detailed description of the detector setup and the running conditions have been given in ~\cite{nanova, Elsner}.
\begin{figure*} 
 \resizebox{0.9\textwidth}{!}{
     \includegraphics[height=.35\textheight]{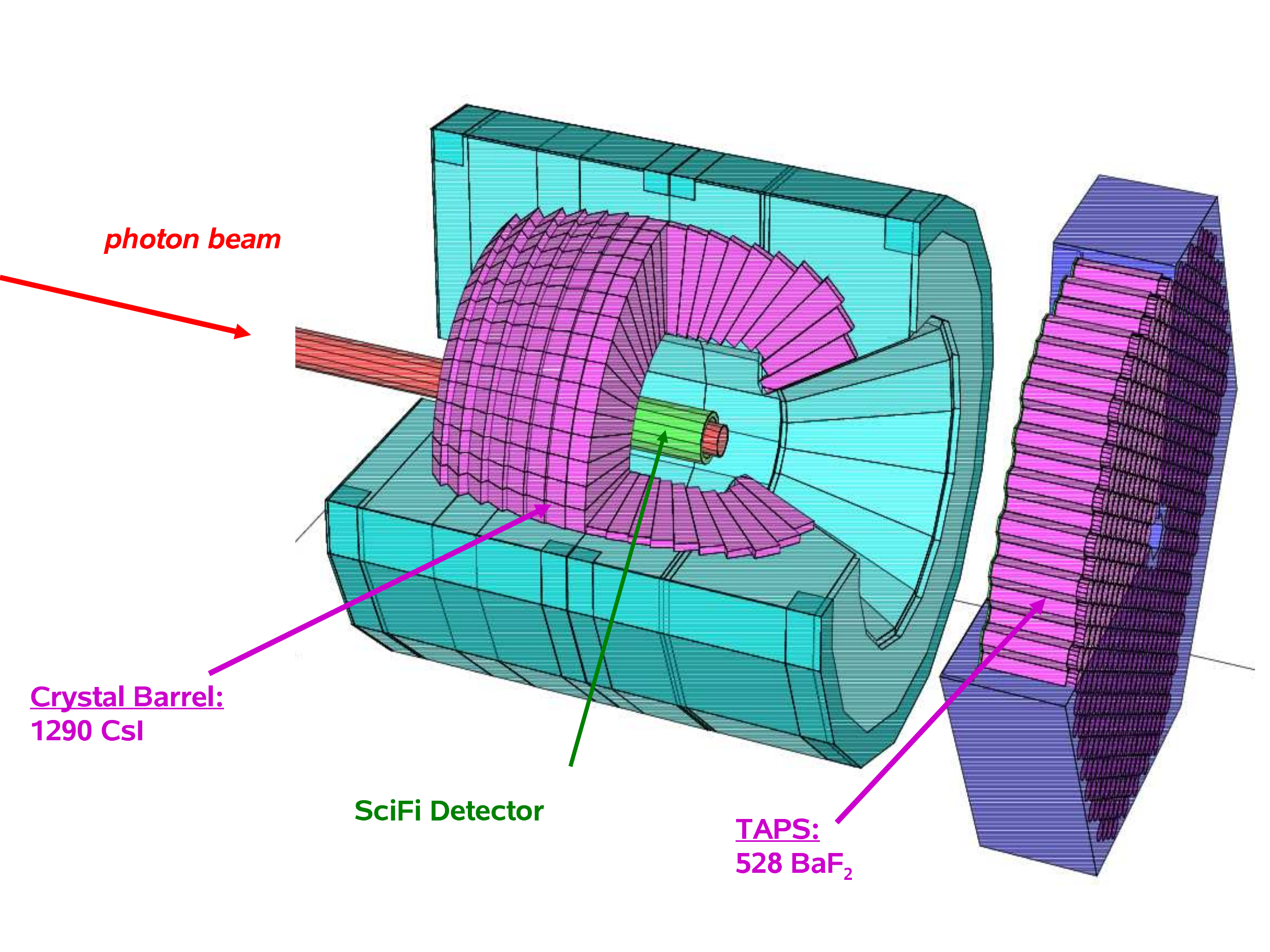}
      \includegraphics[height=.35\textheight]{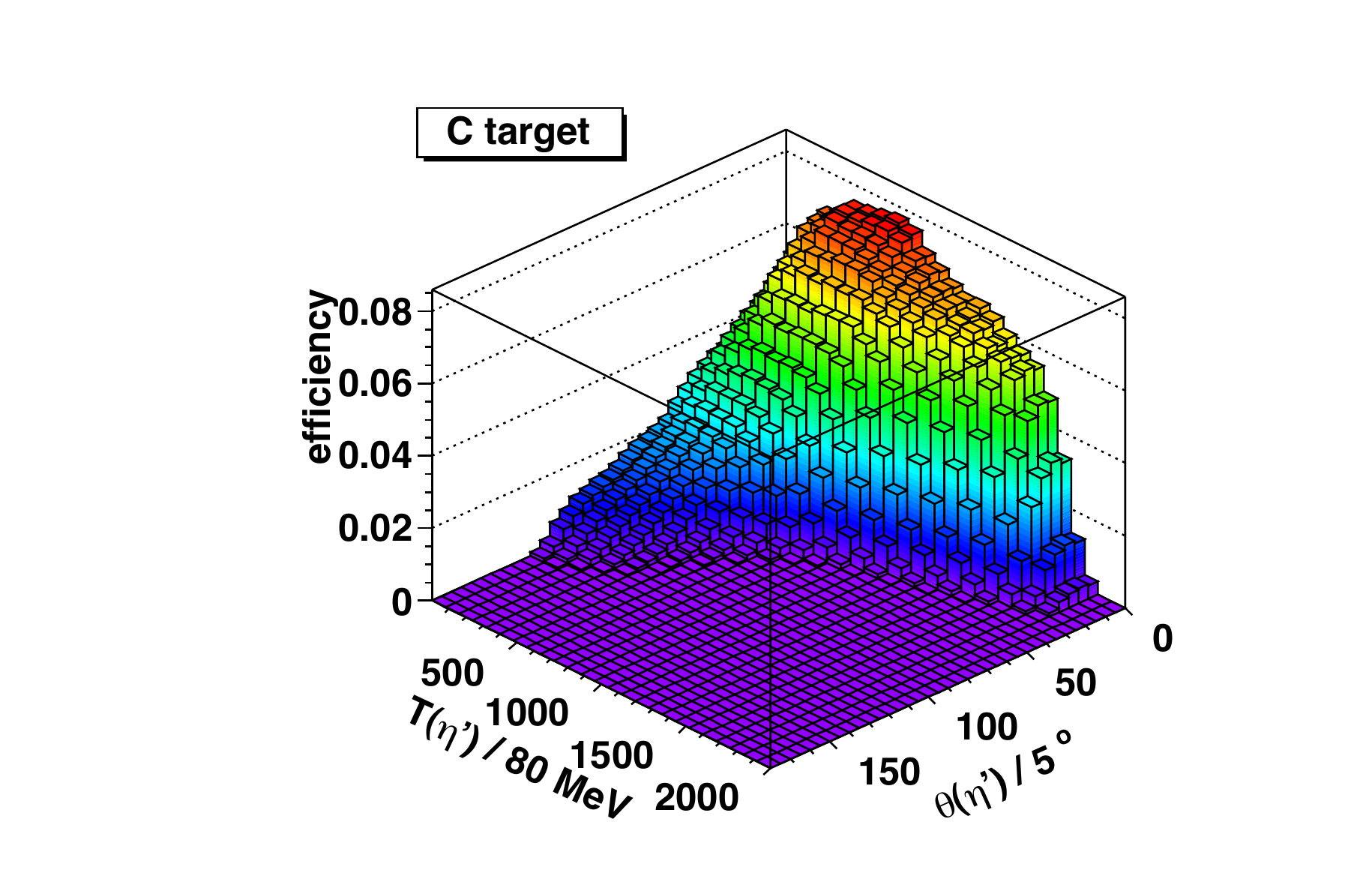}}
 \caption{(Color online) Left: Sketch of the  CB/TAPS setup. The tagged photons impinge on the nuclear target in the center of the Crystal Barrel detector. The   TAPS detector at a distance of 1.18 m from the target serves as a forward wall of the Crystal Barrel. The combined detector system provides photon detection capability over almost the full solid angle. Charged particles leaving the target are identified in the inner scintillating-fiber detector and in the plastic scintillators in front of each BaF$_2$ crystal in TAPS. Right: Detector acceptance for the $ \eta^{'}\rightarrow \pi^{0} \pi^{0} \eta$ decay as a function of the kinetic energy (T) and  the polar angle ($\theta$)  for an  incident photon energy range of 1200 to 2200 MeV. The simulation is for a carbon ($C$) target, taking the trigger conditions into account.} \label{fig:exp} 
\end{figure*}

The detector acceptance was determined by Monte Carlo simulations using the GEANT3 package, including all features of the detector system, trigger conditions and all cuts for particle indentification. To avoid further uncertainties due to reaction kinematics and final state interactions, the $\eta^{'}$-meson detection efficiency was simulated as a function  of the kinetic energy and the polar angle - $\epsilon(T_{\eta'}, \theta_{\eta'})$. Typical efficiencies are 7\% and slighly different for the different targets. The detection efficiency for $\eta'$ on the $C$ target, taking the trigger conditions into account, is shown in Fig. \ref {fig:exp} (right). Experimental data are efficiency corrected event-by-event with this acceptance as described in ~\cite{igal}.\\

The $\eta^{'}$-mesons were identified via the $\eta^{'} \rightarrow \pi^{0} \pi^{0} \eta \rightarrow 6 \gamma$ decay channel, which has a branching ratio of 8\%. For the reconstruction of the $\eta^{'}$-mesons only events with at least 6 or 7 neutral hits have been selected. The competing channel with the same final state, namely $\eta \rightarrow \pi^{0} \pi^{0} \pi^{0} \rightarrow 6 \gamma$,  has been reconstructed and the corresponding events have been rejected from the further analysis. In addition only events were kept with at least one combination of the 6 photons to two photon pairs with invariant masses between 110 and 160 MeV ($\pi^{0}$) and one pair between 500 and 600 MeV ($\eta$). The $\pi^{0} \pi^{0} \eta$ invariant mass distributions for the different solid targets are shown in Fig.~\ref{fig:invmass}. The resulting cross sections are used to calculate the transparency ratio of the $\eta^{'}$-meson for a given nucleus $A$ from the formula~(\ref{eq:trans}), normalized to the carbon data:
\begin{equation}
T_A=\frac{12 \cdot \sigma_{\gamma A\to \eta^\prime A^\prime}}{A \cdot
\sigma_{\gamma C\to \eta^\prime C} } \ .
\label{eq:trans}
\end{equation}

 \begin{figure}[htb]
 \begin{center}
    \includegraphics[width=0.35\textwidth]{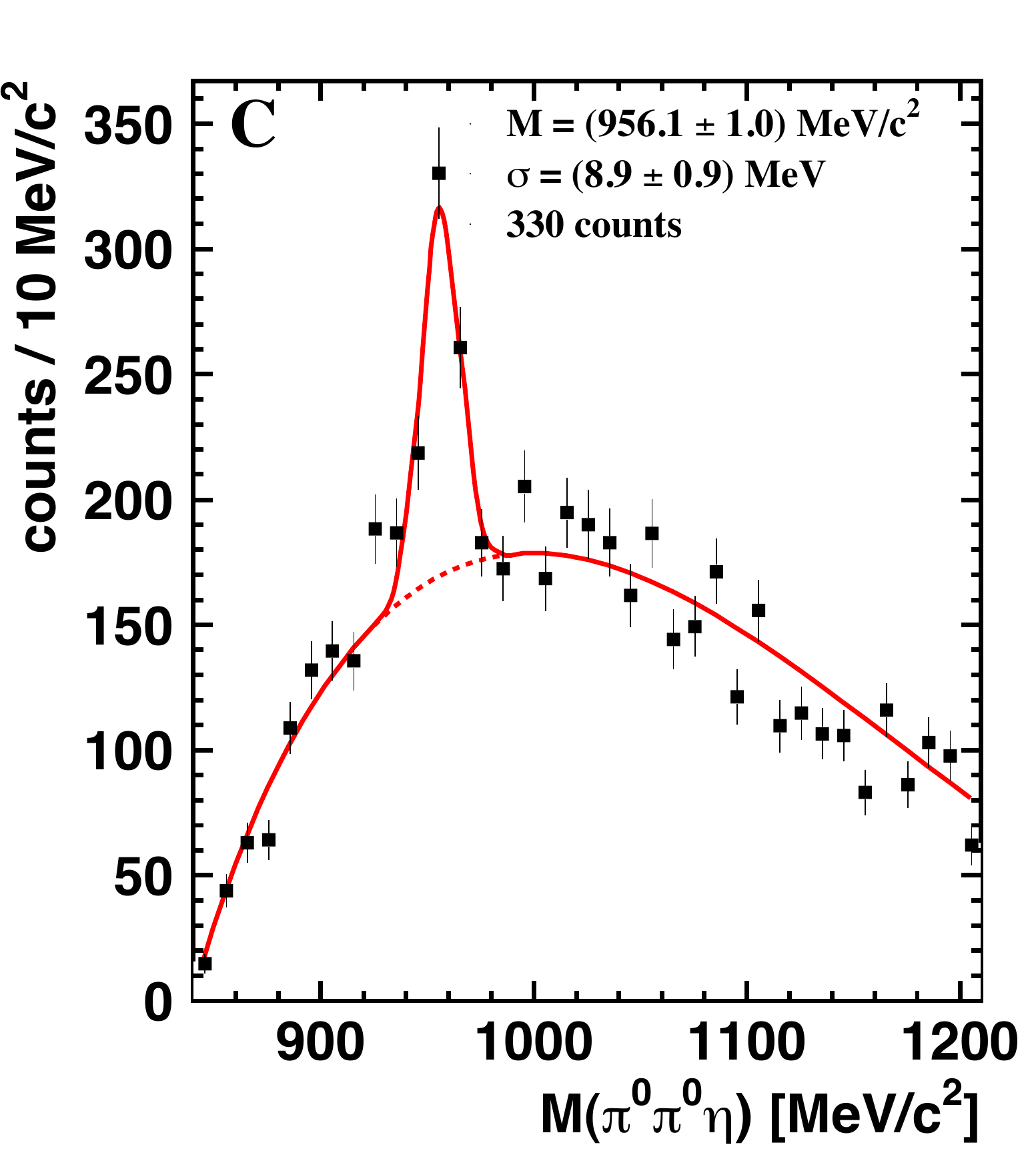}\includegraphics[width=0.35\textwidth]{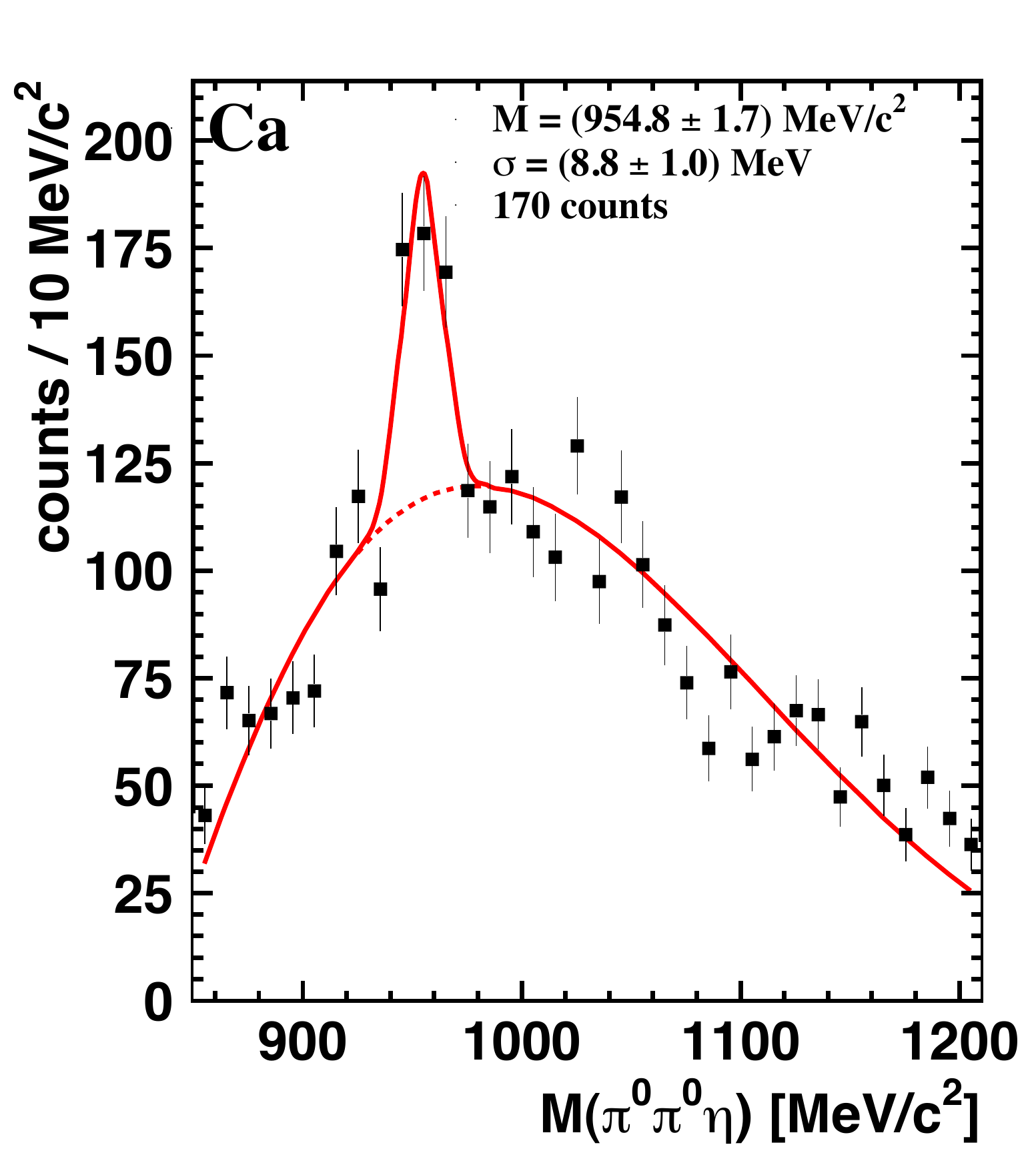}
    \includegraphics[width=0.35\textwidth]{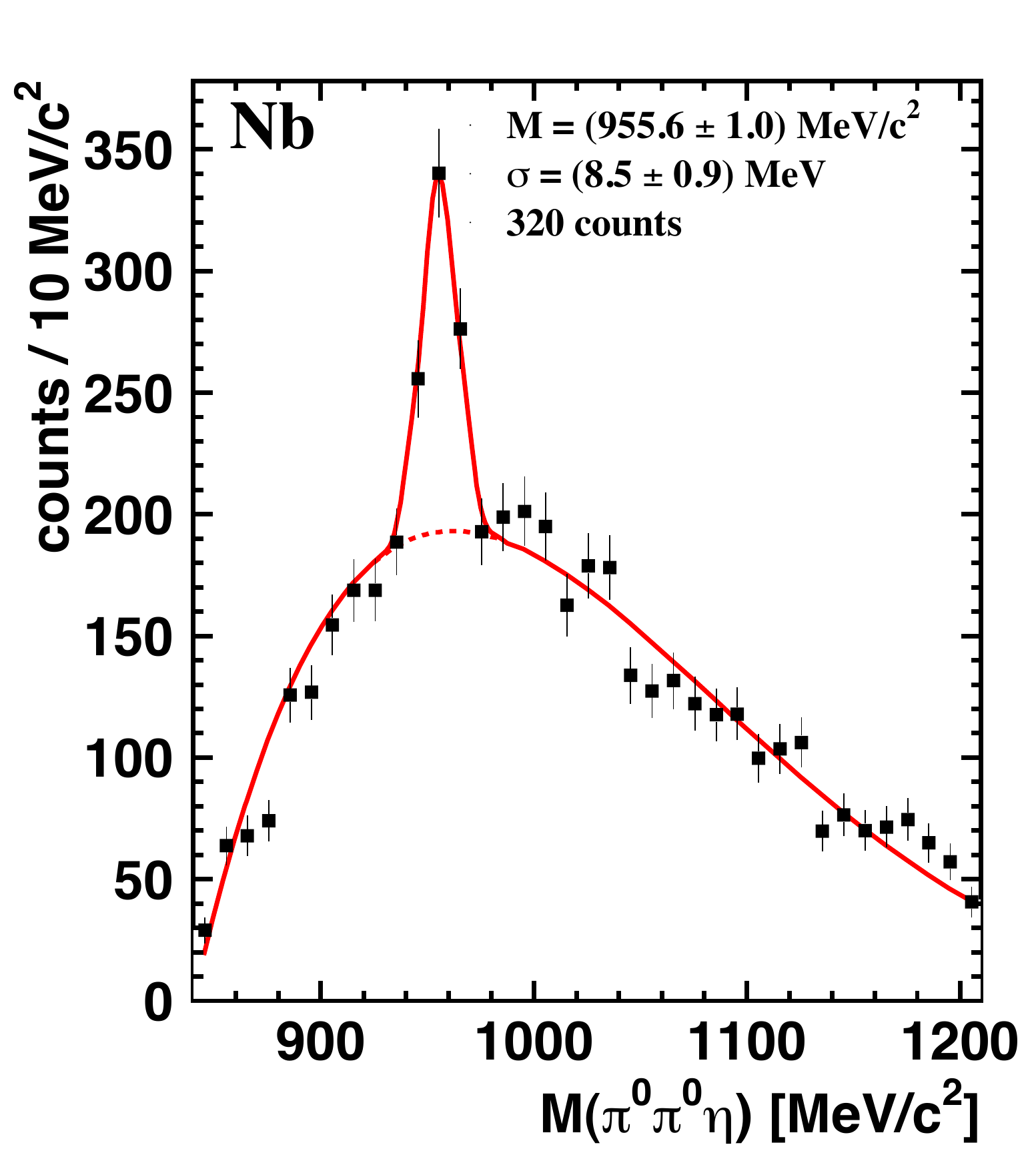}\includegraphics[width=0.35\textwidth]{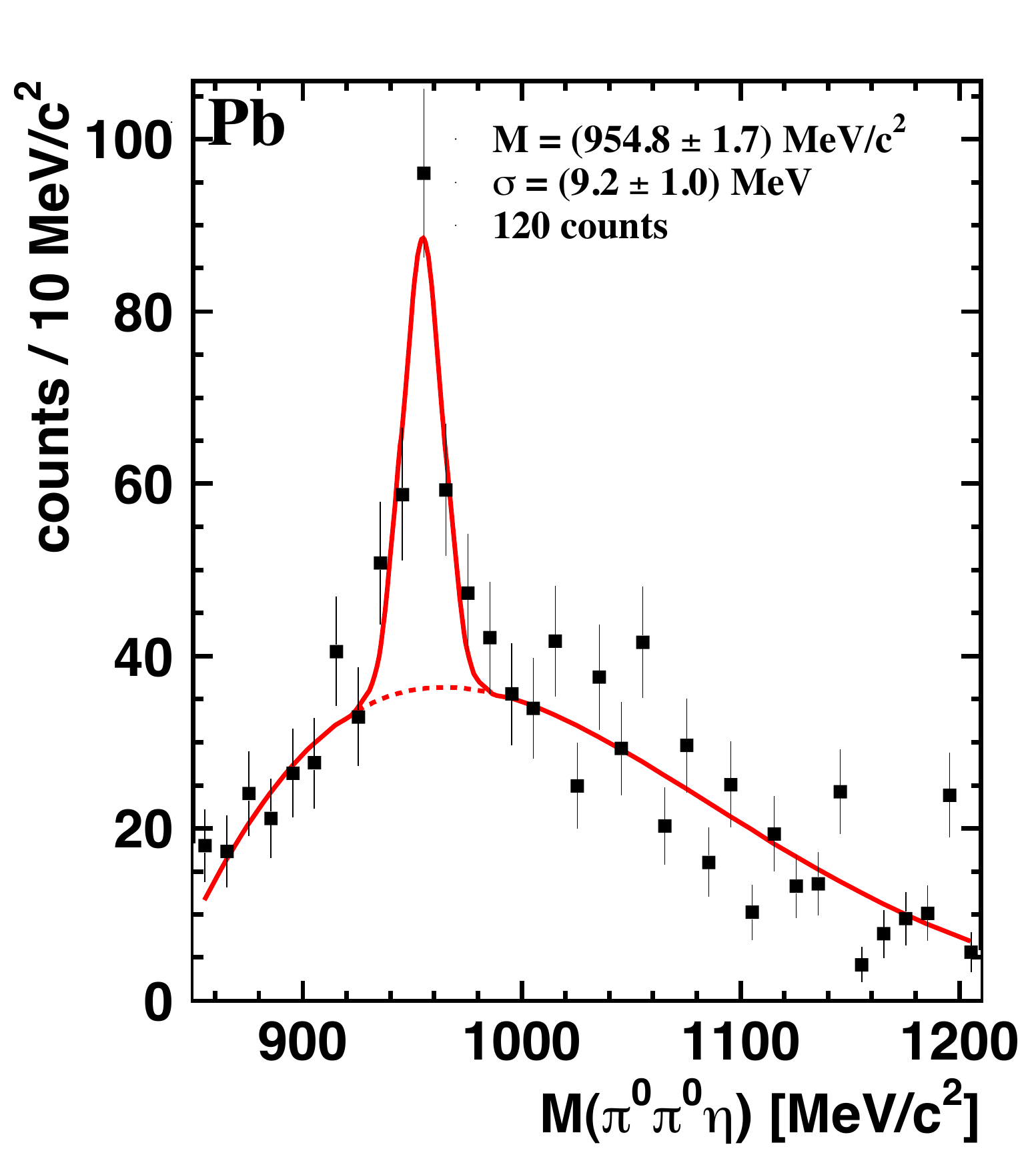}
\caption{Invariant mass spectrum of $\pi^{0}\pi^{0}\eta$ for ${}^{12}\textrm{C}, {}^{40}\textrm{Ca}, {}^{93}\textrm{Nb}$ and ${}^{208}\textrm{Pb}$ targets in incident photon energy range 1200 - 2200 MeV. The distributions have not been corrected for the detector acceptance. The solid curve is a fit to the spectrum. See text for more details.}
\label{fig:invmass}
 \end{center}
\end{figure}

$T_{A}$ describes the loss of flux of $\eta'$-mesons in nuclei via inelastic processes like: $\eta' N \rightarrow \pi^{0} N$. To avoid systematic uncertainties due to unknown secondary production processes, the transparency ratio has been normalized to a light target with equal numbers of protons and neutrons (C) and not to the cross section on the nucleon.

\section{Results and Discussion}
The transparency ratio has been extracted as defined in Eq.~\ref{eq:trans} and is shown in Fig.~\ref{fig:ta} as full (red) triangles. The data are compared to the transparency ratio of the $\omega$-meson measured by~\cite{kottu}. The solid lines are fits to the data points, yielding slope parameters of -0.14 and -0.33 for $\eta^{'}$ and $\omega$, respectively.   As it can be seen from the figure, $\eta^{'}$-mesons are only weakly absorbed via inelastic channels like $\eta^{'} N \rightarrow \pi N$ as compared to the $\omega$-meson. Information on the in-medium width of the $\eta^{'}$ is still not available since theoretical calculations are needed. It is important to study also the contribution of secondary production processes like $\pi N \rightarrow  \eta^{'} N$, which could increase the number of observed $\eta{'}$-mesons and thus distort the transparency ratio. 

 \begin{figure}[htb]
 \begin{center}
    \includegraphics[width=0.5\textwidth]{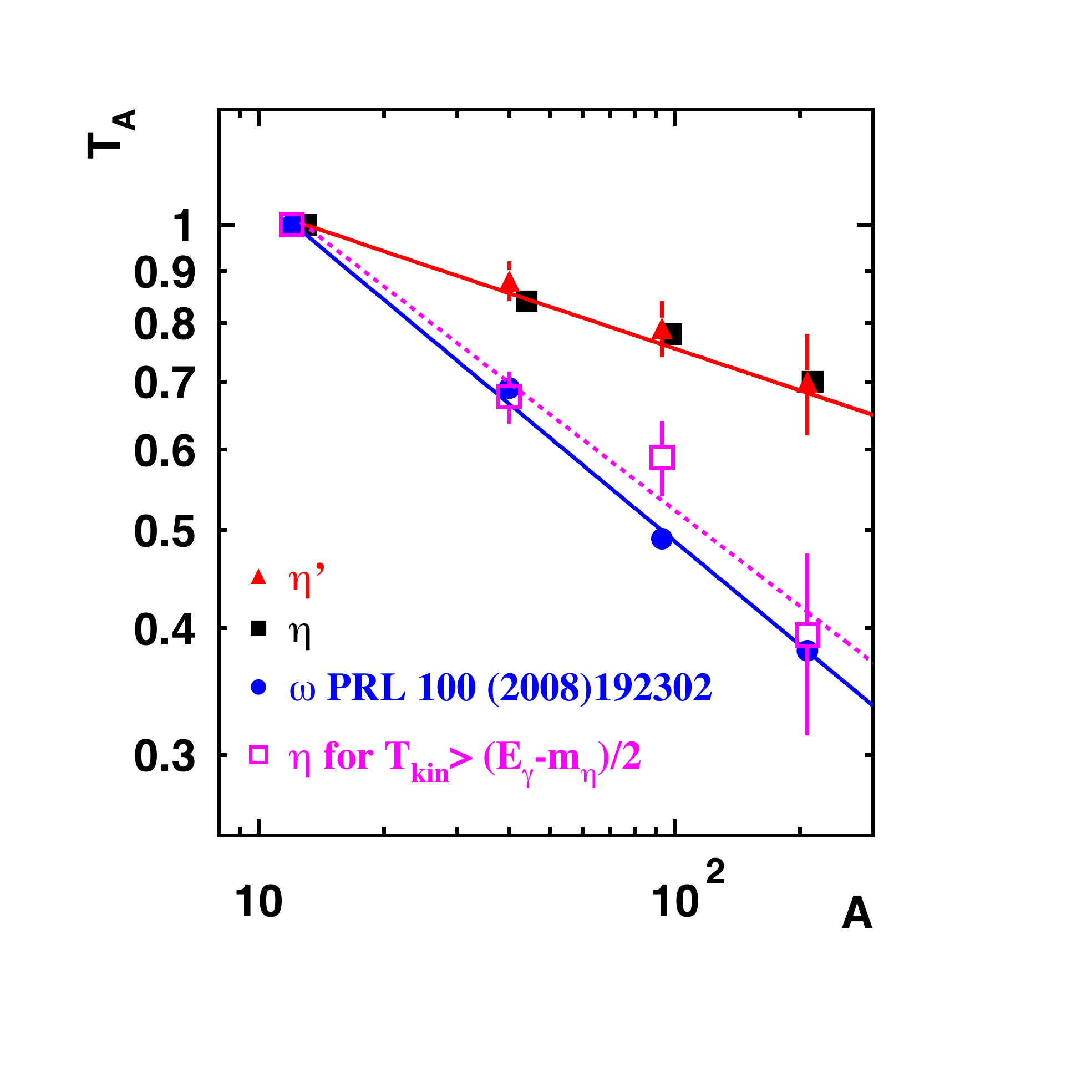}
\caption{Transparency ratio for the $\eta'$-meson (full red triangles) in the incident photon energy range 1200 - 2200 MeV compared to the transparency ratio for the $\omega$-meson (full blue circles) ~\cite{kottu} and for the $\eta$-meson without (full black sqares) and with (empty magenta squares) a cut on the kinetic energy. The solid curve is a fit to the spectrum. See text for more details.}
\label{fig:ta}
 \end{center}
\end{figure}

At least for $\eta$ photoproduction this effect has been studied and  has been found to have a  significant contribution ~\cite{thierry}. As it can be seen in Fig.~\ref{fig:ta}, after applying the condition to select $\eta$-mesons with higher kinetic energy as suggested in~\cite{thierry} the transparency ratio has changed dramatically and shows a slope of -0.33 which is quite different from the previous one - the empty (magenta) squares in Fig.~\ref{fig:ta}. For this reason a new analysis of the data has been started with a condition to select only $\eta^{'}$-mesons with a higher kinetic energy to suppress the contribution from $\pi N \rightarrow \eta^{'} N$ channels. The results will be prepared for publication very soon.

\acknowledgements{%
  We thank the scientific and technical staff at ELSA and the collaborating institutions for their important contribution to the experiment. This work is supported by DFG through SFB/TR 16 "subnuclear structure of matter" and by the Schweizerischer Nationalfond. 
}


%

}  



\begin{thebibliography}{99}
\bibitem{nagahiro}
H.~Nagahiro, M.~Takizawa, S.~Hirenzaki, Phys. Rev. C, {\bf 74}, 045203 (2006).

\bibitem{crede}
V.~Crede {\it et al.},~CBELSA/TAPS Collaboration, Phys. Rev. C, {\bf 80}, 055202 (2009).

\bibitem{igal}
I.~Jaegle {\it et al.},~CBELSA/TAPS Collaboration, Eur. Phys. J. A {\bf 47}, 11 (2011).

 \bibitem{Husmann}
D.~Husmann, and W.~J. Schwille, Phys. Bl. {\bf 44}, 40 (1988).

\bibitem{Hillert}
W.~Hillert,  Eur. Phys. J. A {\bf 28}, 139 (2006).

 \bibitem{aker92}
 E.~Aker {\it et al.}, Nucl. Instr. and Methods A {\bf 321}, 69 (1992).

\bibitem{Novotny1}
R.~Novotny {\it et al.}, IEEE Trans. Nucl. Sci. {\bf 38}, 392 (1991).

\bibitem{nanova}
M.~Nanova {\it et al.},~CBELSA/TAPS Collaboration, Phys. Rev. C {\bf 82}, 035209 (2010).

 \bibitem{Elsner}
D.~Elsner {\it et al.}, Eur. Phys. J. A {\bf33}, 147 (2007).
\bibitem{kottu}
M.~Kotulla {\it et al.}, Phys. Rev. Lett. {\bf 100}, 192302 (2008).
\bibitem{thierry}
T.~Mertens {\it et al.},~CBELSA/TAPS Collaboration, Eur. Phys. J. A {\bf 38}, 195 (2008).

\end{thebibliography}
\end{document}